\documentclass[aps,prb,twocolumn,10pt,superscriptaddress]{revtex4-2}
\usepackage{header}
\usepackage{adjustbox}
\usepackage{physics}
\usepackage{tikz}
\usetikzlibrary{shapes.geometric, arrows.meta, positioning}
\newtheorem{theorem}{Theorem}[section]

\newtheorem{protocol}[theorem]{Protocol}

\newcommand{\ketb}[1]{\ket{\mathbf{#1}}}

\begin{document}
\title{
Comparing Schemes for Creating Qudit Graph States from 16- \& 128-dimensional Hilbert Space using Donors in Silicon
}
\author{G\"ozde \"Ust\"un}
\affiliation{School of Electrical Engineering and Telecommunications, UNSW Sydney, Sydney,
NSW 2052, Australia}
\affiliation{ARC Centre of Excellence for Quantum Computation and Communication Technology, Melbourne, VIC, Australia}
\affiliation{Centre for Quantum Software and Information, University of Technology Sydney, Sydney, New South Wales 2007, Australia}
\author{Simon J. Devitt}
\affiliation{Centre for Quantum Software and Information, University of Technology Sydney, Sydney, New South Wales 2007, Australia}
\affiliation{InstituteQ, Aalto University, 02150 Espoo, Finland}
\begin{abstract}
In this work, we compare two schemes for generating arbitrary qudit graph states using spin qudits in silicon. The first scheme proposes the creation of qudit linear graph states from a single emitter—a silicon spin qudit. By employing fusion—a destructive and non-deterministic measurement technique—these linear graphs can then be combined to form more complex resource states (multi-photon entangled states), such as ring or ladder structures, which are used to carry out the computation.
The second scheme employs two spin qudits. Instead of relying on fusion, the two emitters are directly coupled via CZ to generate the same resource states, thereby eliminating the need for fusion. 
We compare the two schemes in terms of their ability to produce equivalent resource states and discuss their respective advantages and limitations for building scalable architectures. 
\end{abstract}
\maketitle
\section{Introduction}


Fusion-Based Quantum Computing (FBQC) is a promising model for scalable photonic quantum computation, relying on two key physical ingredients: resource states and fusion operations. Resource states are multi-photon entangled states that are measured to carry out the computation. 
While some FBQC protocols can operate with relatively simple resource states like GHZ or linear cluster states~\cite{Paseni_2023}, others may require more complex graph states to support fault tolerance and efficient computation~\cite{psiQ}.

In the qubit regime, one well-known example of resource states is the 6-ring graph state, which has been studied extensively due to its compatibility with surface code implementations~\cite{bombin_logical,bombin2023faulttolerantcomplexes}. While not inherently optimal for photon-loss tolerance, encoded versions of the 6-ring have been shown to improve robustness. More generally, increasing the complexity of resource states — such as adding redundancy or encoding — can enhance loss tolerance, as explored in recent studies~\cite{PRXQuantum_paseni}.
These states can be created in two main ways: (1) Multiplexing, which relies on combining heralded single photons from nondeterministic sources (e.g. silicon waveguides, cavities, or nonlinear crystals) to approximate deterministic generation. However, this comes at the cost of a very high resource overhead due to the repeated execution of these nondeterministic operations~\cite{Lindner_2009} and the main challenge is the complex routing required for multiplexing, which involves many optical elements (such as switches and detectors). These components are all lossy, limiting the ability of an integrated device to achieve the high level of reliability (or probability of success) needed for fault-tolerant applications. 
(2) Quantum emitters can, in principle, generate graph states deterministically and with far greater resource efficiency. However, in practice this method faces hardware‐level challenges, such as photon distinguishability issues, and fabrication imperfections~\cite{Huet_2025}. 
In this approach, the excitation of a quantum emitter, typically a semiconductor quantum dot, and the subsequent emission of a photon through relaxation behaves like a CZ gate, creating entanglement between consecutively emitted photons~\cite{Hilaire2023,Pettersson2025,Chan2025,Lobl2025}. There are two main strategies here:

\begin{itemize}
     \item Single emitter combined with fusion operations~\cite{Paeseni_emitter_3,Huet_2025}. A linear graph state can be created deterministically from a single quantum emitter, without any form of multiplexing~\cite{Lindner_2009}. Fusion operations, which are non-deterministic, destructive measurements, can then be used to construct more complex arbitrary shapes beyond simple linear graphs. 
     \item  Another version involves using multiple quantum emitters and creating entanglement between them~\cite{sophia_1, sophia_2, Raissi_2024}. This approach does not require fusion within the same graph and allows for the direct, deterministic creation of arbitrarily complex graphs, by coupling the emitters via CZ gates~\cite{Raissi_2024}. The fusion is only necessary between multiple (different) resource states to carry out the computation.
\end{itemize}
By integrating resource states with fusion operations, we fulfill the two essential physical requirements of FBQC. Both elements — resource states and fusion operations — have been extensively explored in the qubit regime~\cite{bartolucci2021fusionbasedquantumcomputation, bharos2024efficienthighdimensionalentangledstate, bartolucci2021creationentangledphotonicstates, Paeseni_Emitter, Paeseni_emitter_2, Paeseni_emitter_3, sophia_1, sophia_2}.

Photons however naturally serve as qudits, as they can occupy multiple modes and deterministically form high-dimensional states. For a $d$-dimensional linear-optical qudit, arbitrary single-qudit unitaries can be implemented deterministically using only linear optical components. When paired with a single entangling operation, these tools enable the creation of arbitrary multi-qudit states, thereby supporting universal high-dimensional quantum computation~\cite{Paesani, Lib_2024}.

Qudit graph states generalize qubit graph states by introducing weighted edges expressed as powers of CZ gates~\cite{helwig2013absolutelymaximallyentangledqudit}, in contrast to the qubit case where each edge between emitted photons corresponds to a single CZ gate. Their deterministic generation is feasible using qudit quantum emitters, presenting promising opportunities in both qubit and qudit regimes~\cite{Lib_2024,Raissi_2024}. While photon loss poses a challenge, quantum error correction offers a solution. In essence, using qudits allows more information to be stored with the same level of error resilience, without increasing the number of physical systems. The $[[5,1,3]]_{\mathbb{Z}_d}$ qudit code~\cite{chau1997five} and the qudit surface code~\cite{bullock2007qudit} are two such examples. More precisely, in~\cite{Watson_2015}, it was shown that the threshold of the qudit surface code increases with the dimension, compared to the standard surface code. For example, for qudit dimension 8, the threshold increases to above 3\%.
Moreover, recent advances in fusion operations within high-dimensional linear optical systems~\cite{bharos2024efficienthighdimensionalentangledstate, Ustun_fusion}, combined with qudit graph states, pave the way for scalable fusion-based quantum computing in higher dimensions—together forming the essential physical ingredients of FBQC.

In this work, we present a comparative study from the hardware perspective on the creation of qudit cluster states based on antimony donors — high-spin donors in silicon. First, we propose generating linear qudit graph states from a single antimony donor and fusing photons within the chain to construct more complex qudit resource states. Second, we consider a system of two antimony donors hyperfine-coupled to the same electron and propose generating similar resource states via direct coupling. We compare these two approaches and discuss their respective experimental advantages and limitations. 

The structure of the paper is as follows. In Section~\ref{sec:pre}, we briefly define the necessary concepts such as qubit and qudit graph states as well as antimony donor. In Section~\ref{single_ant}, we present the hardware, Hamiltonian, key components, and the protocol for creating linear qudit graph states using a single antimony donor. In Section~\ref{fusion}, we demonstrate the fusion of linear qudit graph states and show how to form ring structures, along with an analysis of their success probabilities. In Section~\ref{double_ant}, we introduce our hardware setup consisting of two antimony nuclei hyperfine-coupled to a shared electron, and present protocols for generating arbitrary graph state shapes without requiring fusion operations within the same graph. Finally, in Section~\ref{comparison}, we compare the two hardware approaches, discussing their respective advantages and disadvantages.

\section{PRELIMINARIES}\label{sec:pre}
Note: If the reader is familiar with the concept of graph states and  (high-dimensional) donors in silicon, they may proceed directly to Section~\ref{single_ant}.

\subsection{Recap of Graph States}

A quantum graph state is defined by a graph \( G = (V, E) \), where \( V \) is the set of vertices and \( E \) is the set of edges that specify the entanglement relationships among the vertices. Each vertex \( v \in V \) corresponds to a photon initialized in the state \( \ket{+} \). For every edge \( (u, v) \in E \), a controlled-Z (CZ) gate is applied between the qubits associated with vertices \( u \) and \( v \), thereby generating entanglement.

\begin{figure}[H]
\centering
\begin{minipage}[t]{0.49\textwidth}
    \raggedright
    \textbf{(a)}\\
    \includegraphics[width=\linewidth]{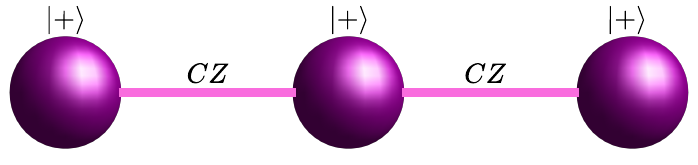}
  \end{minipage}
  \hfill
  \begin{minipage}[t]{0.44\textwidth}
    \raggedright
    \textbf{(b)}\\
    \centering
    \includegraphics[width=.7\linewidth]{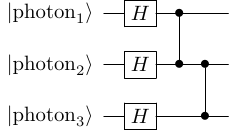}
  \end{minipage}
  \caption[Three-photon graphs state]{a) represents a three-photon graph state where the purple spheres correspond to photons prepared in the $\ket{+}$ state. The photons are connected by edges, which correspond to CZ gates. b) shows the circuit representation of the graph state defined in (a). A Hadamard gate is applied to each photon all of which are initialized in the state $\ket{0}$, followed by CZ gates between the first and second photons, as well as between the second and third photons, forming a linear cluster state. Here, photons are treated as qubits; they are not yet qudits.}
  \label{fig:three_cluster}
\end{figure}

The Figure~\ref{fig:three_cluster} illustrates a three-qubit linear graph states where a $CZ$ gate is applied between the first and second photons, as well as between the second and third photons. The resultant state of the circuit in Figure~\ref{fig:three_cluster} is 
\[
\ket{000} + \ket{001} + \ket{010} - \ket{011} + \ket{100} + \ket{101} - \ket{110} + \ket{111}
\] up to a normalization coefficient. 

A qudit graph state can then be constructed by replacing each operation with its qudit counterpart. In this case, the Hadamard operation becomes the qudit Hadamard, or, as defined in Section~\ref{ingredients}, it can be written as the Fourier gate $F_d$. The CZ gate is also replaced by its qudit counterpart, which can be written as
$
\text{CZ}_{ij} = \sum_{k=0}^{d-1} \ket{k}\bra{k}_i \otimes Z_j^{k},
$
where $Z$ is the generalized Pauli-$Z$ operator for a qudit, defined as $Z\ket{k} = \omega^k \ket{k}$, with $\omega = e^{2\pi i/d}$ being the $d$-th root of unity. A detailed explanation of qudit graph states can be found in~\cite{helwig2013absolutelymaximallyentangledqudit}. Very briefly, we can define them as follows:

An $n$-qudit graph, with each qudit having dimension $d$, is represented by a set of vertices $V = \{v_i\}$ connected by edges $E = \{e_{ij} = \{v_i, v_j\}\}$. Each edge is assigned a weight $A_{ij} \in \mathbb{Z}_d$, where a weight of zero corresponds to the absence of an edge. The weights $A_{ij}$ form a symmetric $n \times n$ adjacency matrix with $A_{ii} = 0$, which captures all relevant information about the graph~\cite{helwig2013absolutelymaximallyentangledqudit}.  

For a given graph $G$ with $n$ vertices and adjacency matrix $A \in \mathbb{Z}_d^{n \times n}$, 
the corresponding graph state $|G\rangle \in \mathcal{H}^{\otimes n}$, with $\mathcal{H} \cong \mathbb{C}^p$, is defined as
\begin{equation}
|G\rangle = \prod_{i>j} \text{CZ}_{ij}^{A_{ij}} \, |\bar{+}\rangle^{\otimes n}. \label{eq:graph_state}
\end{equation}


A graph state can be constructed using a quantum circuit that first prepares all qudits in the $\ket{\bar{+}}$ state, which is typically a superposition of the computational basis states. For example, in the qutrit case, $\ket{\bar{+}} = \ket{0} + \ket{1} + \ket{2}$, assuming that the eigenstates of the $Z$ operator are defined as $\ket{0}$, $\ket{1}$, and $\ket{2}$. Pairwise CZ gates are then applied according to the entries of the adjacency matrix. Hence, in the qutrit case, there exist two CZ gates corresponding to $A_{i,j} = 1$ and $A_{i,j} = 2$.

\subsection{Antimony Donor}

We propose a CMOS-compatible silicon-based quantum architecture using antimony (\(^{123}\)Sb) donors with a nuclear spin of \(I = 7/2\). This spin state alone defines an 8-dimensional Hilbert space, calculated as \(2I + 1 = 8\). As a group V element with five valence electrons, \(^{123}\)Sb behaves like a hydrogenic impurity, capable of binding an additional electron at cryogenic temperatures (see Fig.~\ref{fig:antimony_donor}). The inclusion of this bound electron expands the total Hilbert space to 16 dimensions~\cite{fernandez2024navigating}.

A single antimony atom can be embedded into the silicon lattice via ion implantation, replacing a native silicon atom. The isotope \(^{123}\)Sb features a gyromagnetic ratio \(\gamma_n = 5.55\,\text{MHz/T}\). Its non-spherical nuclear charge distribution gives rise to an electric quadrupole moment in the range \(q_n = [-0.49, -0.69] \times 10^{-28} \, \text{m}^2\). The donor-bound electron, with spin \(S = 1/2\), has a gyromagnetic ratio of approximately \(\gamma_e = 27.97\,\text{GHz/T}\), and couples to the nuclear spin via the Fermi contact hyperfine interaction, described by \(A \hat{S} \cdot \hat{I}\), where \(A = 101.52\,\text{MHz}\) in bulk silicon.

Among group V donor candidates with high nuclear spin, such as \(^{209}\)Bi and \(^{75}\)As, \(^{123}\)Sb presents several advantages. Although bismuth offers a higher nuclear spin (\(I = 9/2\)), its larger atomic mass leads to increased lattice damage and reduced activation efficiency. Additionally, bismuth’s significantly stronger hyperfine coupling (\(A = 1475.4\,\text{MHz}\)) can cause pronounced measurement back-action on the nuclear spin during readout~\cite{joecker2024error}. Arsenic, by contrast, has a lower nuclear spin ($I=3/2$), which restricts the creation of qudit graph states to a local dimension of 4.

Antimony, in particular, is a well-characterized physical system: its Hamiltonian has been determined with precision down to a few hertz, and its noise profile and environmental interactions are thoroughly understood~\cite{asaad2020}. The associated error rates are remarkably low, as demonstrated in our previous experimental studies~\cite{fernandez2024navigating, joecker2024error, asaad2020}.

Moreover, silicon donors host a loosely bound electron, which enables the simultaneous emission of multiple photons when scaling beyond a single antimony atom. For instance, two antimony donors—each with its own bound electron—can emit two photons simultaneously. Extending this to \(N\) donors allows for the concurrent emission of \(N\) photons, facilitating scalable quantum photonic architectures.
\subsection{Hamiltonian of single antimony donor}

\begin{figure}[htb]
    \centering
    \includegraphics[width=.8\linewidth]{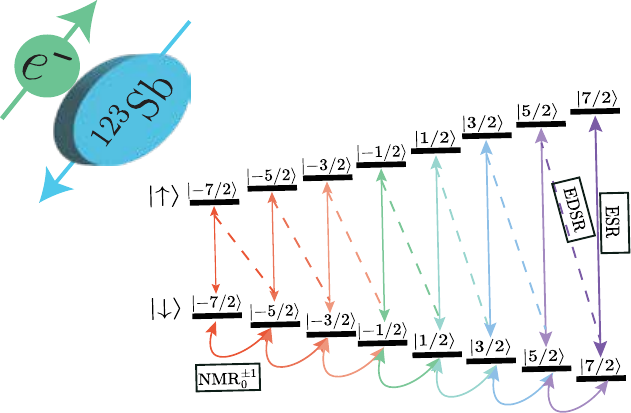}
    \caption{Energy spectrum of the single neutral antimony donor. Antimony possesses a high nuclear spin, giving rise to eight distinct nuclear spin states. In its neutral charge configuration, the electron spin becomes hyperfine-coupled to the antimony nuclear spin, resulting in a Hilbert space of dimension 16. NMR transitions with $\Delta m =\pm 1$ are represented by curved arrows and are labeled with a subscript 0 to denote the neutral charge state. ESR transitions are shown as vertical solid arrows, while EDSR transitions are indicated with dashed arrows.}
    \label{fig:antimony_donor}
\end{figure}
As a group V donor, an antimony (Sb) atom embedded in silicon forms a Coulomb potential well capable of binding a single electron. Together with this electron, the Hamiltonian of the Sb system can be written as follows:
\begin{equation}
    H_{\text{Sb}} = 
         B_0(-\gamma_n \hat{I_z} + \gamma_e \hat{S_z}) + A(\Vec{S}\cdot\Vec{I}) + \Sigma_{\alpha, \beta \in \{x,y,z\}} Q_{\alpha \beta}\hat{I}_\alpha \hat{I}_\beta     
\end{equation}
The details of this Hamiltonian can be found in various previous work of us~\cite{fernandez2024navigating,Yu2025,Ustun2025,Ustun_fusion}. Very briefly, the spin Hamiltonian of the antimony atom built from the tensor product of the electron spin operator and two nuclear spin operators \cite{fernandez2024navigating}. The first term in $H_{\text{Sb}}$ accounts for the Zeeman splitting, on both the electron and the nuclei. The second term describes the hyperfine interaction that arises from the overlap of electron and nuclear wavefunctions. In the last term, where $\alpha, \beta = {x, y, z}$ represent Cartesian axes, $\hat{I}_\alpha$ and $\hat{I}_\beta$ are the corresponding 8-dimensional nuclear spin projection operators. 
The term $Q_{\alpha \beta}$ is the interaction energy between the electric quadrupole moment of the nucleus and the electric field gradient, which arises primarily from local strain that breaks the cubic symmetry of the silicon lattice~\cite{asaad2020coherent}. This quadrupole interaction introduces an orientation-dependent energy shift to the nuclear Zeeman levels, enabling the individual addressability of nuclear states. $B_0$ represents the magnetic field in which the nanoelectronic device containing the antimony donor is placed, with a value approximately equal to $1 \text{T}$.
This ensures that the eigenstates of $H_{\text{Sb}}$ are approximately the tensor products of the nuclear states $\ket{m_I}$ with the eigenstates $ \{\ket{\downarrow} , \ket{\uparrow} \} $ of $\hat{S}_z$ because $\gamma_e B_0 \gg A \gg Q_{\alpha\beta}$. The latter condition implies $H_{\text{Sb}} \approx B_0(-\gamma_n \hat{I_z} + \gamma_e \hat{S_z}) + A(\Vec{S} \cdot \Vec{I})$ ensuring that the nuclear spin operator approximately commutes with the electron-nuclear interaction.
This condition allows for an approximate quantum non-demolition (QND) readout of the nuclear spin via the electron spin ancilla \cite{joecker2024error}.

Our proposed device features a silicon nanoelectronic component fabricated on top, as described in~\cite{Ustun2025,Ustun_fusion}. This component includes a broadband microwave antenna that delivers the $B_1$ field required for coherent transitions. 
Coherent transitions between the $^{123}$Sb spin eigenstates can be induced by magnetic and electric fields, on both the electron and the nuclear spin~\cite{fernandez2024navigating}. Electron spin resonance (ESR), which flips the spin of the electron, is achieved by the driving term $H^{\text{ESR}} = B_1 \gamma_e \hat{S}_x \cos(2\pi f_{m_I}^{\text{ESR}}t) $. Here, $B_1$ is the amplitude of an oscillating magnetic field at one of the eight resonance frequencies $f_{m_I}^{\text{ESR}}$ determined by the nuclear spin projection $m_I$. Nuclear magnetic resonance (NMR), which changes the nuclear spin projection by one quantum of angular momentum, is achieved by the driving term $H^{\text{NMR}} = B_1 \gamma_n \hat{I}_x \mathrm{cos}(2\pi f_{m_I - 1 \leftrightarrow m_I}^{\text{NMR}}t)$. In addition to magnetically driven transitions (ESR, NMR), there exists the possibility of inducing spin transitions electrically. This is accomplished by leveraging the combined states of the electron and nucleus, representing a high-spin extension of the `flip-flop' transition recently demonstrated in the $I = 1/2$ $^{31}$P system~\cite{Rosty}. The application of an oscillating electric field $E_1 \cos\left(2\pi f_{m_I-1\leftrightarrow m_I}^{\text{EDSR}} t\right)$ leads to electric dipole spin resonance transitions (EDSR) in the neutral donor. This process dynamically modulates the hyperfine interaction $A(E_1) \hat{S}_{\pm} \hat{I}_{\mp}$ through the Stark effect~\cite{tosi_2017}, where the $\pm$ subscripts denote the raising and lowering operators, respectively. This mechanism conserves the total angular momentum of the combined electron-nuclear states. Consequently, the EDSR transitions manifest as diagonal (dashed) lines in Fig~\ref{fig:antimony_donor}.
\section{Linear Qudit Graph States from Single Antimony donor}\label{single_ant}
We begin by explaining our proposed system: an antimony donor coupled with a microwave cavity, along with its Hamiltonian. Next, we outline the necessary operations to implement the protocol for creating linear qudit graph states. Finally, we demonstrate how the fusion operation can be performed to form a ring structure.

\subsection{Proposed System}
We propose integrating microwave cavities into the device, enabling the coherent emission of single photons via a time-bin multiplexing scheme, as shown in~\cite{Ustun2025}

\begin{figure}[htb]
    \centering
    \includegraphics[width=\linewidth]{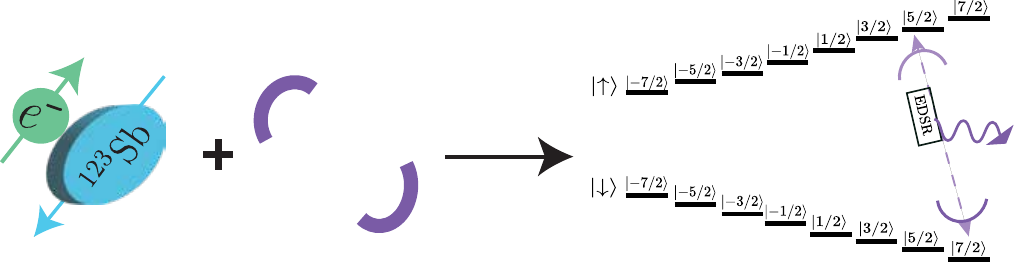}
    \caption{The antimony donor is assumed positioned at an antinode of the cavity
electric field, to achieve capacitive coupling to the donor’s electric dipole and stimulate the EDSR transitions to emit photons coherently. The EDSR frequency between the states $\ket{7/2}\ket{\downarrow} \leftrightarrow \ket{5/2}\ket{\uparrow}$ is chosen as fixed frequency for emitted photons}
    \label{fig:antimony_cavity}
\end{figure}
The total Hamiltonian of the system is given by:
\begin{equation}
    H_{\text{total}} = H_{Sb} + H_{\text{interaction}} + H_{\text{field}}
\end{equation}

$ H_{\text{interaction}} + H_{\text{field}}$ part represents the emitting photon through cavity. The total Hamiltonian can be expressed as follows:
\begin{equation}\label{drift_Ham}
    \begin{aligned}
        H_{\text{total}} = & H_{\text{Sb}}
         + (\underbrace{ \hbar w_{\ket{7/2}\leftrightarrow\ket{5/2}} a^{\dag} a}_{H_{\text{field}}}       
        +\\ &   \underbrace{g (\ket{7/2}\ket{\downarrow}\bra{5/2}\bra{\uparrow} a^{\dag}) + 
         g(\ket{5/2}\ket{\uparrow}\bra{7/2}\bra{\downarrow} a)}_{H_{\text{interaction}}} )\\       
    \end{aligned}
\end{equation}
 where the operators $a^{\dag}$ and $a$ represent creation and annihilation operators, respectively. $g$ represents the coupling strength between the spin and the cavity. When purely magnetic coupling is employed via the ESR transitions, $g$ is in the range of $10$Hz - $100$kHz~\cite{Wyatt_1}. However, when utilizing an electric dipole for spin-cavity coupling, significantly higher coupling strengths -- in the range of a few MHz\cite{tosi_2017} -- and thus faster photon emission times are possible. This is why we propose to operate the cavity at the antimony donor EDSR transitions.

\subsection{Ingredients for protocols}\label{ingredients}
We now list the necessary operation for our protocols:
\begin{itemize}
    \item A permutation gate which permutes the population in between different nuclear spin states. This can be achieved with a single NMR pulse ($\pi$ pulse on the nucleus or equivalently nuclear $X$-gate) for nearest-neighbor states such as $\ket{7/2}$ and $\ket{5/2}$ or $\ket{5/2}$ and $\ket{3/2}$, and so on. For non-neighboring states (such as $\ket{7/2}$ and $\ket{-7/2}$), sequential NMR pulses or more sophisticated control schemes, such as global rotations described in~\cite{Yu2025}, can accomplish the desired transitions. In the case of sequential NMR, leakage can be a problem. In contrast, advanced control methods like global rotations, which permute the intermediate states between any two desired states, can offer more reliable performance for transitions between distant states. However, the required timing may exceed that of consecutive NMR pulses.
    \item A $d$-dimensional Fourier gate, which is equivalent to the qudit Hadamard gate for creating a uniform superposition of chosen nuclear states, is defined by the following formula:
    \begin{equation}    
    F_d = \frac{1}{\sqrt{d}} \sum_{j,k=0}^{d-1} \omega^{jk}\ket{j}\bra{k} 
     \end{equation} 
where $\omega = e^{2\pi i/d}$ which is the dth root of unity. When $d=2$, the fourier gate is simply qubit hadamard gate.
    \item An ESR pulse is used to transfer the population of a nuclear state to the electron's spin-up state from its spin-down state, and vice versa. An EDSR pulse can also be used to address the corresponding anti-diagonal spin states.
    \item A coherent interaction between spin of the electron and the microwave cavity
    \item Measuring the antimony donor. This is a simple experimental process and done by sequentially applying an ESR pulse conditional on the nuclear states, and flipping the spin state of the electron from $\ket{\downarrow}$ to $\ket{\uparrow}$. Then, readout the electron via spin to charge conversion. This fundamental process is performed on many other experiments See details in ~\cite{Morello_2010, Stemp2024, Fern_ndez_de_Fuentes_2024, Yu2025}.
\end{itemize}.
\subsection{The protocol}
We use a time-bin multiplexing scenario to create qudit graph states. We first describe how to emit a single photon into multiple dimensions, and then show how to create linear qudit graph states from a single antimony donor.

\begin{figure*} 
    \includegraphics[width=\linewidth]{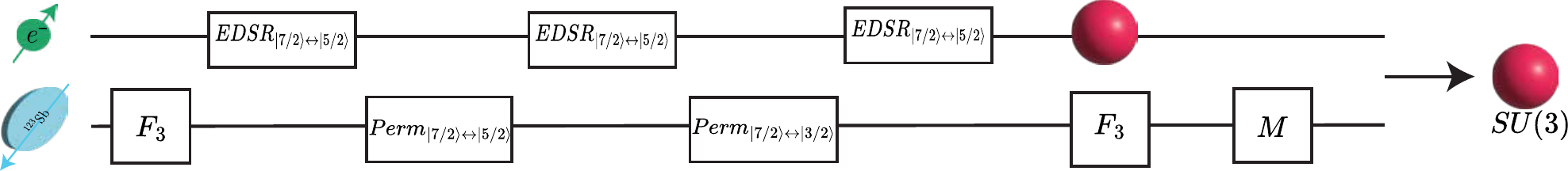}
    \caption{The circuit illustrates the steps for emitting single photons using the time-bin multiplexing method, in which we generate a three-dimensional single photon from a single antimony donor. We begin by applying a generalized Hadamard gate ($F_3$) on the nucleus. Next, we apply an EDSR pulse at the frequency at which the cavity operates (between the states $\ket{7/2}$ and $\ket{5/2}$) on the electron to excite the population of state $\ket{7/2}\ket{\downarrow}$ to $\ket{5/2}\ket{\uparrow}$, allowing the electron to emit a photon coherently. We then apply a permutation operation between the $\ket{7/2}$ and $\ket{5/2}$ states to exchange the populations between states. In this case, the permutation is achieved via a single nuclear magnetic resonance (NMR) pulse, followed by another EDSR pulse. This process of applying a permutation and an EDSR pulse is repeated sequentially until the third state involved in the superposition, $\ket{3/2}$, is used. After these operations, a single photon is emitted in 3 modes but remains entangled with the antimony donor. To decouple the antimony from the photon, we apply a second $F_3$ gate and then measure the antimony. As a result, a single photon is emitted into many modes. \label{fig:photon_emission}
}
\end{figure*}

Note: A more detailed version of this part of the protocol is provided in~\cite{Ustun2025}, where a photon is encoded into eight time-bins. For completeness, we will outline the protocol again, but this time for a three-dimensional photon, in which case the photon serves as a qutrit. However, with antimony and time-bin multiplexing, it is possible to encode a photon into more than three dimensions.

Let us now begin describing the protocol for emitting a single photon into multiple modes. Since we will present a small example for the case \( d = 3 \), three energy levels of antimony are required instead of two. We begin with the initial state, \( \ket{7/2}\ket{\downarrow}\ket{\text{vac}} \), and then create a uniform superposition of the three energy levels by applying a qudit Hadamard gate, in this case $F_3$. The state then becomes:
\[
    \begin{split}
        \ket{\psi_1} = \frac{1}{\sqrt{3}}\left( \ket{7/2} + \ket{5/2} + \ket{3/2} \right) \ket{\downarrow} \ket{\text{vac}}.
    \end{split}
\]

We use an EDSR pulse to flip the spin of the electron 
conditional on the nuclear state being \( \ket{7/2} \). The electron then undergoes a coherent exchange of energy with the microwave cavity and returns to the spin-down state, while the cavity becomes populated with a photon of frequency $\omega$ at $t_1$
\[
    \begin{split}
        \ket{\psi_2} = \frac{1}{\sqrt{3}} \ket{7/2}\ket{\downarrow} \ket{\omega}_{t_1}  + \left( \ket{5/2} + \ket{3/2} \right) \ket{\downarrow} \ket{\text{vac}}.
    \end{split}
\]

Next, we apply a permutation operation between \( \ket{7/2} \) and \( \ket{5/2} \):

\[
    \begin{split}
        \ket{\psi_3} = \frac{1}{\sqrt{3}} \ket{5/2}\ket{\downarrow} \ket{\omega}_{t_1}  + \left( \ket{7/2} + \ket{3/2} \right) \ket{\downarrow} \ket{\text{vac}}.
    \end{split}
\]
After permuting the populations, we then repeat the excitation process for the state $\ket{7/2}$, followed by a permutation operation between \( \ket{7/2} \) and \( \ket{3/2} \):
\[
    \begin{split}
        \ket{\psi_4} = \frac{1}{\sqrt{3}} & ( \ket{5/2}\ket{\downarrow} \ket{\omega}_{t_1}\ket{vac}  +  \\ &\ket{3/2} \ket{\downarrow} \ket{\text{vac}}\ket{w}_{t_2} + \\ & \ket{7/2}\ket{\downarrow}\ket{vac}\ket{vac}).
    \end{split}
\]
We finally apply the last excitation and emission step for the remaining state.
The resulting state becomes:
\[
    \begin{split}
        \ket{\psi_f} = \frac{1}{\sqrt{3}}\ket{\downarrow}(& \ket{7/2} \ket{vac}\ket{vac}\ket{\omega}_{t_3}  + \\  &\ket{5/2} \ket{\omega}_{t_1}\ket{vac}\ket{vac}  + \\&\ket{3/2}\ket{vac}\ket{\omega}_{t_2})\ket{vac}.
    \end{split}
\]
Applying a qudit Hadamard gate followed by a measurement of the antimony nuclear spin decouples the nuclear state from the photonic modes. Expressing the final state in the binary-encoded Fock basis yields:
\[
    \begin{split}
        \ket{\psi_f} = &\frac{1}{3} \left( \ket{7/2} + \ket{5/2} + \ket{3/2} \right) \ket{\downarrow} \ket{001} \\
        &+ \left( \ket{7/2} - \ket{5/2} - \ket{3/2} \right) \ket{\downarrow} \ket{100} \\
        &+ \left( \ket{7/2} - \ket{5/2} + \ket{3/2} \right) \ket{\downarrow} \ket{010}.
    \end{split}
\]

A single photon is emitted into three time-bins and the resultant state is essentially equivalent to a three-dimensional \( W \) state (\( \ket{W_3} \)). Upon measurement of the antimony nuclear spin, the photonic state collapses to \( \ket{100} + \ket{010} + \ket{001} \), up to a state specific phase operation.

Figure~\ref{fig:photon_emission} shows the circuit visualization of what we just described above. It is now clear how this approach generalizes to higher dimensions: to increase the qudit dimension from three to $n$, we need to use $n$ energy levels instead of the three energy levels of antimony. For example, when eight energy levels of antimony are used, the resulting state becomes a $\ket{W_8}$ state instead of $\ket{W_3}$.

To create a linear qudit graph state from a single antimony donor, instead of measuring the donor, we proceed by exciting the spin and then emitting photons via the microwave cavity. 
The protocol can be written as follows:
\onecolumngrid
\begin{protocol}\label{pro:linear_qudit_graph}
Creation of $d$-dimensional $n$-photon linear qudit graph states from a single antimony donor via a time-bin multiplexing protocol. Note, the steps highlighted in pink represent the remaining steps for emitting a single photon into $d$ time bins after the qudit Hadamard gate is applied.

\begin{center}
\begin{tikzpicture}[
    node distance=8mm,
    every node/.style={font=\small},
    process/.style={
        rectangle,
        rounded corners,
        draw,
        text width=8.2cm,
        align=center,
        minimum height=6mm
    },
    decision/.style={
        diamond,
        draw,
        aspect=2,
        text width=4cm,
        align=center,
        inner sep=1pt
    },
    arrow/.style={->, thick}
]

\node (step1) [process] {Choose $d$ antimony nuclear-spin levels and the number of graph nodes $n$.};

\node (step2) [process, below=of step1] {Initialize the system in
$\ket{\psi_i} = \ket{7/2}\ket{\downarrow}\ket{\mathrm{vac}}$.};

\node (step3) [process, below=of step2] {Apply a qudit Fourier (Hadamard) gate across the selected $d$ nuclear-spin states.};

\node (step4) [process, below=of step3, fill=pink] {Apply an EDSR pulse to flip the electron spin conditioned on the nuclear spin being in $\ket{7/2}$.};

\node (step5) [process, below=of step4, fill=pink] {Let the electron interact coherently with the cavity, emitting a photon of frequency $\omega$.};

\node (decision1) [decision, below=of step5, yshift=-2mm, fill=pink] {Are all $d$ energy levels of antimony used in the superposition for the required $d$-dimensional single-photon?};

\node (step6) [process, right=20mm of decision1, fill=pink] {Permute $\ket{7/2}$ with the next unused nuclear-spin state.};

\node (step8) [process, below=10mm of decision1] {Apply a qudit Hadamard gate to each nuclear-spin state.};

\node (decision2) [decision, below=of step8, yshift=-2mm] {$n$ photons generated?};

\node (step9) [process, right=20mm of decision2] {Repeat the emission sequence to generate the next node in the linear qudit graph.};

\node (step10) [process, below=10mm of decision2] {Measure the antimony donor after a final qudit Hadamard gate to decouple nuclear and photonic states.};

\draw [arrow] (step1) -- (step2);
\draw [arrow] (step2) -- (step3);
\draw [arrow] (step3) -- (step4);
\draw [arrow] (step4) -- (step5);
\draw [arrow] (step5) -- (decision1);

\draw [arrow] (decision1.east) -- node[above]{No} (step6.west);
\draw [arrow] (step6.north) |- (step4.east);

\draw [arrow] (decision1.south) -- node[right]{Yes} (step8.north);

\draw [arrow] (step8) -- (decision2);

\draw [arrow] (decision2.east) -- node[above]{No} (step9.west);
\draw [arrow] (step9.north) |- (step4.east);

\draw [arrow] (decision2.south) -- node[right]{Yes} (step10.north);

\end{tikzpicture}
\end{center}
\end{protocol}

\twocolumngrid
\subsection{Fusion of Linear Qudit Graph States}\label{fusion}

\begin{figure*}[ht]
    \centering
    \includegraphics[width=.8\linewidth]{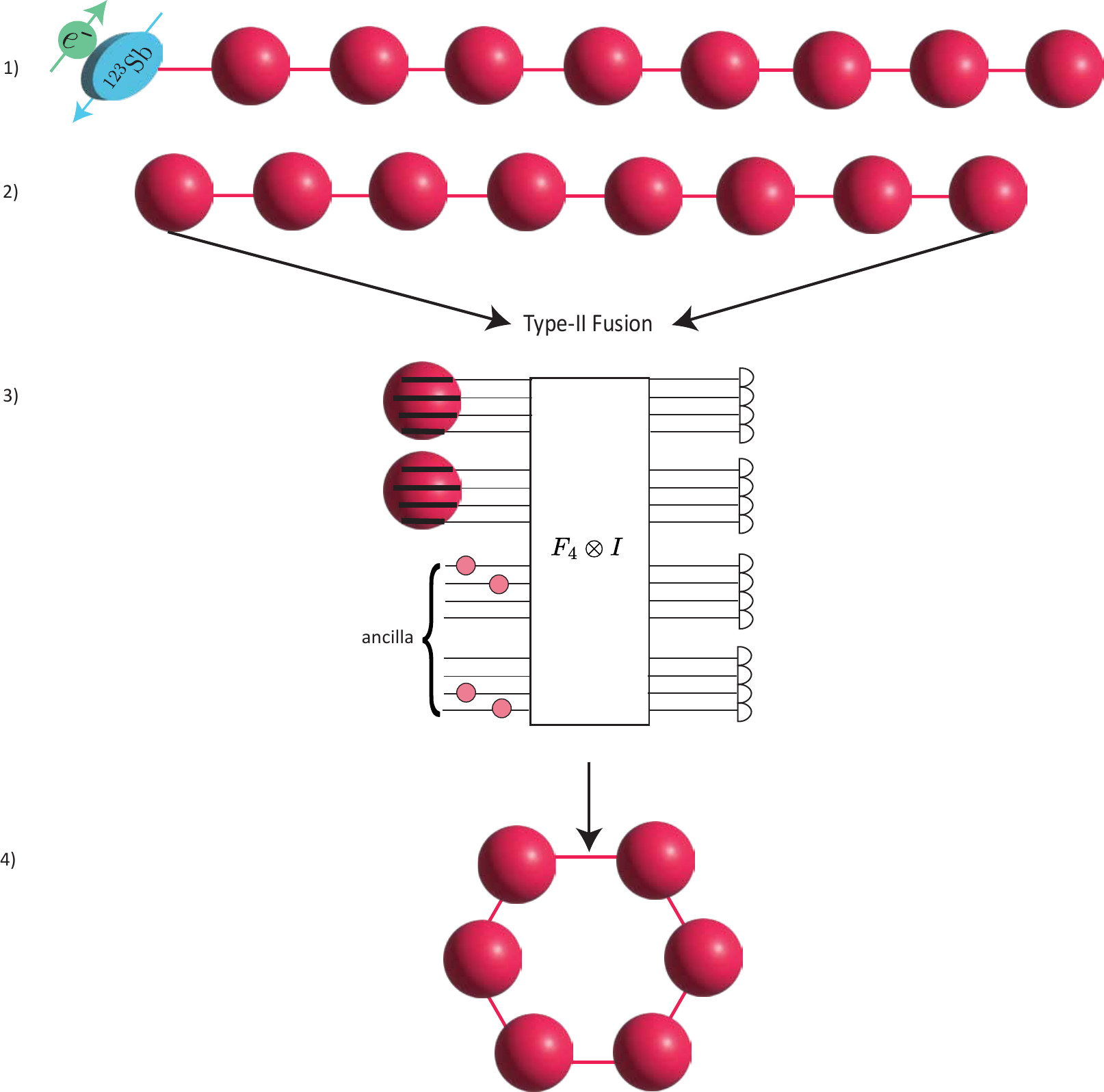}
    \caption{\textbf{Type-II Fusion of a Linear Eight-Qudit (Four-Dimensional) graph State.} This figure illustrates the generation of high-dimensional resource states for photonic quantum computing. In 1), an eight-qudit linear graph state is prepared, with local dimension, qudit dimension being four. In 2), the antimony donor is then measured to decouple the nuclear spin from the photons, yielding a purely photonic graph state. 3) Subsequently, the first and eighth photons are fused using a type‑II fusion gate, which requires an entangled ancilla state consisting of a superposition of two photons across eight modes. 4) This operation results in a six-ring qudit graph state, where each node (qudit) retains its four-mode encoding.}
    \label{fig:fusion}
\end{figure*}
In the case of linear graphs, resource states can be engineered either by creating a single extended linear graph state and fusing its first and last high-dimensional photons (Figure~\ref{fig:fusion}), or by fusing two independent linear graph chains to form a high-dimensional graph state of arbitrary shape, depending on where the fusion is performed. Recent advances have improved the success probabilities of fusion protocols for high-dimensional graph states, enabling fusion for both odd and even dimensions with an approximate success rate of $\sim 2/d^2$~\cite{Ustun_fusion}. By applying such type-II fusion gates to linear graph states, a variety of high-dimensional resource states can be constructed for fusion-based quantum computing. The type-II fusion gate can be interpreted as a $d$-dimensional Fourier operation involving $d$ qudits, each of dimension $d$. The first $d$ modes correspond to the first qudit, the second $d$ modes correspond to the second qudit, and so on. The total number of modes in the state to which the fusion is applied is $d^2$. Each $d$-block can be regarded as a \emph{port}. Therefore, a type-II fusion gate for qudits requires two $d$-mode ports for the fusion, with ancillary states occupying the remaining $d - 2$ ports ($d(d-2)$-modes for the ancilla state). Figure~\ref{fig:fusion} illustrates the best-known type-II fusion protocol which requires an entangled state as an ancilla in the form of:
\begin{equation}
    \ket{A_d} = \frac{1}{\sqrt{d/2}}\sum_{r=0}^{d/2-1} \ket{\mathbf{0+2r}, \mathbf{1+2r}, \dots, \mathbf{d-3+2r}},
\end{equation}
where the indices are taken modulo $d$. ($d-2$ photons across $d-2$ ports, each with $d$ time bins). The superposition will always have $d/2$ terms, and the $r$th term has qubits in all possible computational basis states except for $\ketb{d+2r-2}, \ketb{d+2r-1}$. In the $d=4$ case, the ancilla has $2$ photons in $8$ modes and is given by 
\begin{align}
    \ket{A_4} &= \frac{1}{\sqrt{2}}(\ket{10000100} + \ket{00100001})
    \\&= \frac{1}{\sqrt{2}}(\ket{\mathbf{01}} + \ket{\mathbf{23}}).
\end{align}
We note that, if modes $1,3,4,6$ (all of which are empty) are omitted, the state $\ket{A_4}$ is seen to be simply the \emph{two-dimensional} Bell pair $\frac{1}{\sqrt{2}}(\ket{1010} + \ket{0101})$. 
Another explicit example, for $d=6$ we have
\[\ket{A_6} = \frac{1}{\sqrt{3}}\left( \ketb{0123} + \ketb{2345} + \ketb{4501} \right).\]
The state may be simplified by omitting empty modes again: one can remove the odd-indexed modes from the even-indexed qudits and vice versa. By removing these redundant modes and applying cyclic shifts, we see that $\ket{A_d}$ is equivalent to a $(d-2)$-GHZ state in $\frac{d}{2}$ dimensions. ($\ket{A_4}$ is a $2$-dimensional Bell pair, $\ket{A_6}$ is a $3$-dimensional $4$-GHZ state, etc.) In this instance, for qudit dimensions beyond 3 and 4, the required ancilla state can be constructed using additional antimony donors, as demonstrated in~\cite{Ustun_fusion}. The explicit construction of the antimony donor is described in detail in~\cite{Ustun_fusion}.

The generalization of this protocol to arbitrary dimension $d$ involves applying a Fourier transformation (Fourier gate) along with ancilla states spanning $d - 2$ modes. In Fig.~\ref{fig:fusion}, the Fourier transform is written as $F_d \otimes I$ due to our convention for representing $n$ linear-optical qudits of dimension $d$: the first $d$ modes form the first qudit, the next $d$ modes form the second, and so on. Although the physical implementation may vary, we refer to the $d$ modes within each qudit as ``time bins'' for convenience. It is often helpful to index modes as pairs $(i,j)$ corresponding to mode $d*i + j$, where $j$ is the $j$th time bin of the $i$th qudit. In settings where qudits correspond to spatial ports and their internal modes correspond to time bins, this convention makes the operation equivalent to applying a Fourier interferometer only across spatial modes while leaving the time bins unchanged, hence the notation $F_d \otimes I$.

Given that the antimony donor provides access to a 16-dimensional Hilbert space, it is feasible to construct qudit graph states with local dimensions significantly greater than four. However, the coherence time of the antimony, as well as the operational timescales for constructing graph states—such as the implementation of generalized $d$-dimensional Hadamard gates, application of permutation operations—must be taken into careful consideration.

It is also important to note that, despite recent improvements, the success probabilities of fusion operations remain lower than in the qubit case. For instance, the method introduced in~\cite{Ustun_fusion} achieves a success probability of $2/d(d+1)$ for odd dimensions, which corresponds to probabilities of $0.066$ for $d = 5$ and $0.0357$ for $d = 7$. For even dimensions, the success rates are similarly constrained, with $0.125$ for $d = 4$ and $0.055$ for $d = 6$, all notably lower than those achievable in the qubit regime.


\section{Direct Generation of Arbitrary Graph States from Coupled Quantum Emitters}\label{double_ant}
An alternative to the single-emitter and probabilistic fusion approach is to use coupled emitters, enabling arbitrary graph state generation without fusion, at least within a single graph. We now present a device with two antimony donors sharing one electron to explain how this works.
\subsection{\texorpdfstring{The system: Two Antimony one electron (Sb$_2^+$)}{The system: Two Antimony one electron (Sb2+)}}

\begin{figure}[htb]
   \includegraphics[width=.9\linewidth]{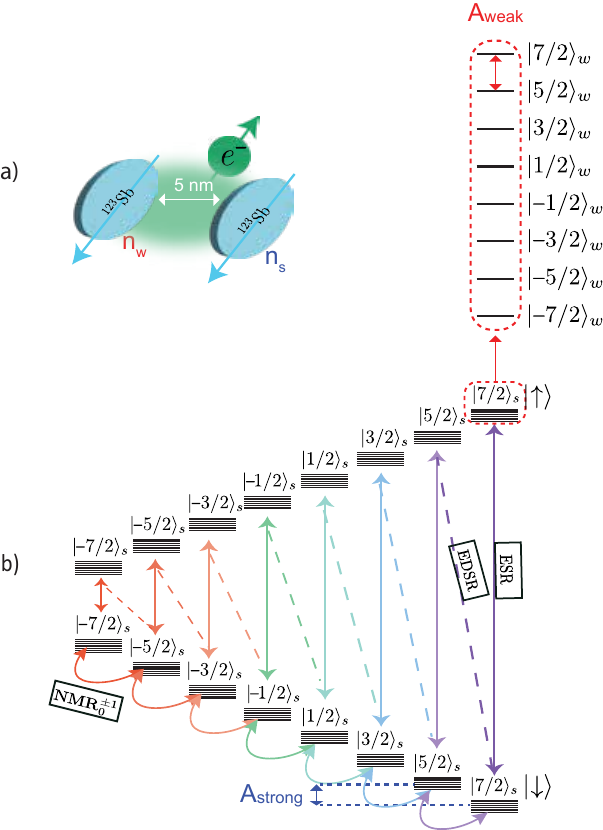}
   \caption{$Sb{_2}^+$ molecule. The molecule under investigation consists of two spin-$\tfrac{7}{2}$ $^{123}$Sb nuclei which are asymmetrically hyperfine-coupled to a single bound electron. The two $^{123}$Sb nuclei are spatially separated by approximately 5~nm. The nucleus exhibiting a stronger hyperfine interaction with the electron is denoted as $n_s$, while the one with a weaker coupling is referred to as $n_w$. b) Spin states energy diagram of the molecule. The Hilbert space spans over $2\times 8 \times 8 = 128$ states. The hyperfine coupling $A_s$ between the electron and $n_s$ makes the ESR frequency strongly dependent on the spin state of the strongly coupled nucleus ($\ket{m_S}$). Every state is then further split in 8 states separated by $A_w $ due to the weak hyperfine coupling between the electron and $n_w$.}
    \label{fig:coupled_antimony}
\end{figure}
We can implant a molecule of antimony ($\mathrm{Sb}_2$), instead of single atom. In the case of an $\mathrm{Sb}_2$ molecule, the two atoms can form a cluster—effectively a single trapping site—where an electron becomes asymmetrically bound to both atoms.

Upon contact with the silicon surface, the molecule dissociates, but the resulting atoms remain in close proximity, typically separated by approximately 5 nm. In this configuration, the electron spin couples asymmetrically to the nuclear spins of the two antimony atoms via the Fermi contact hyperfine interaction—strongly to one nucleus and weakly to the other.

When a static magnetic field $B_0$ is applied along the $z$-axis, the resulting spin system is described by a \textbf{128-dimensional Hamiltonian}. The Hamiltonian of the system is given as follows:
  \begin{equation}
   \begin{aligned}     
    H_{\text{Sb}_{2}^{+}} = &- \gamma_n B_0 (I_z^w + I_z^s) + \\ &\gamma_e B_0 S_z + \\& \hat{S}(A_w I^w + A_s I^s) + \\& \Sigma_{\alpha, \beta \in \{x,y,z\}} Q_{\alpha \beta}^w\hat{I}_\alpha^w\hat{I}_\beta^w + \\ & \Sigma_{\alpha, \beta \in \{x,y,z\}} Q_{\alpha \beta}^s\hat{I}_\alpha^s\hat{I}_\beta^s
    \end{aligned}
  \end{equation}
where upperscripts $w$ and $s$ represent the nuclei coupled with the electron weakly and strongly, respectively. The first term in the Hamiltonian is the Zeeman interaction for both antimony donors. The second term is the Zeeman interaction of the electron. The third term represents the hyperfine coupling, which arises from the overlap between the electron and nuclear wavefunctions. The final term accounts for the quadrupole interaction experienced individually by both nuclei.
The operators $I_\alpha^w$, $I_\alpha^s$, and $S_\alpha$ denote the spin-$7/2$ operators of the two antimony nuclei that are weakly and strongly coupled to the electron, and the spin-$1/2$ operator of the electron, respectively, along the Cartesian axis $\alpha$. The parameters $A_w$ and $A_s$ represent the hyperfine coupling strengths between the electron and the weakly and strongly coupled antimony nuclei, and are equal to 239~kHz and 96~MHz, respectively~\cite{Hsu2025}. The quadrupole interactions $Q_{\alpha \beta}^s$ and $Q_{\alpha \beta}^w$, experienced by the nuclei $n_s$ and $n_w$, arise from the non-spherical charge distribution of the nuclei and are equal to 44.3~kHz and 35.6~kHz, respectively~\cite{Hsu2025}. These interactions couple the nuclear spin states to the electric field gradient generated by electrostatic gates and lattice strain of the silicon chip.


The system comprises two $^{123}$Sb nuclei, each with spin-$7/2$, resulting in $8 \times 8 = 64$ nuclear spin configurations for a given electron spin state. Since the electron can be either spin-up or spin-down, the total Hilbert space dimension is $2 \times 64 = 128$. Therefore, there are 64 allowed ESR and NMR transitions in total. In addition to magnetically driven transitions (ESR, NMR), there exists the possibility of inducing spin transitions electrically. There are 7 allowed EDSR transitions for the strongly coupled donor. Additional EDSR transitions exist for the weakly coupled donor, though these are not shown in the figure. If we restrict our analysis to the subspace in which the nuclear spin state of the strongly coupled donor remains unchanged, then 56 distinct EDSR transitions are available for the weakly coupled donor.

\subsection{\texorpdfstring{Sb$_2^+$ coupled with a Microwave Cavity}{Sb2+ coupled with a Microwave Cavity}}

\begin{figure}[htb]
    \centering
    \includegraphics[width=.9\linewidth]{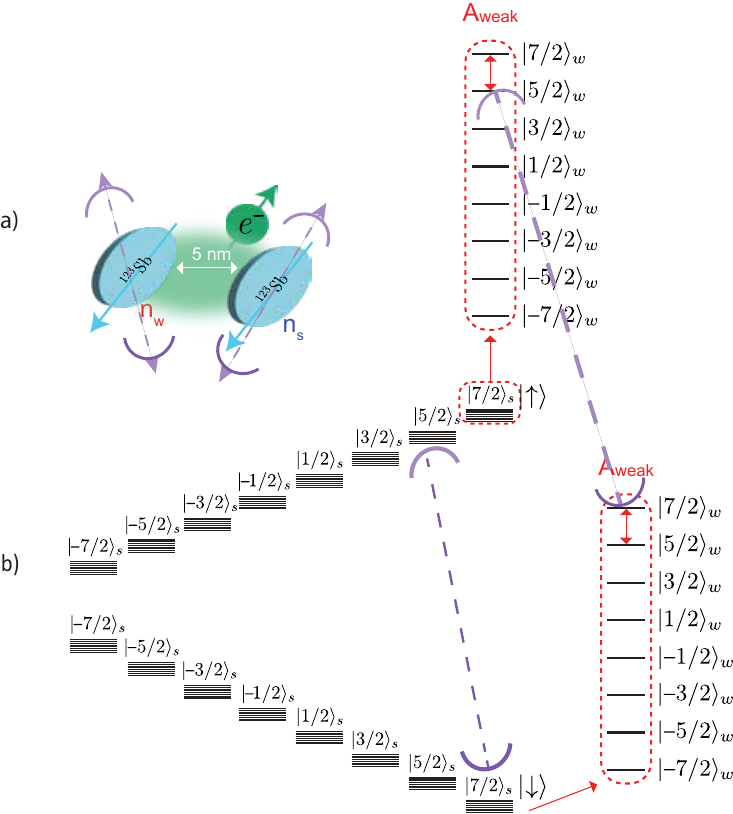}
    \caption{Each antimony donor is individually coupled to a microwave cavity. a) Schematic representation showing each antimony donor coupled to its respective microwave cavity. b) The cavities are tuned to the electric-dipole spin resonance (EDSR) transition frequency between the nuclear spin states $\ket{7/2}$ and $\ket{5/2}$. For the weakly coupled antimony nucleus, this transition is selected such that the strongly coupled nucleus remains in its initial state (i.e., not flipped). It is important to note that the system contains only a single electron, and thus only one cavity can be operated at a time}
    \label{fig:two_cavity}
\end{figure}

Our proposed system comprises two microwave cavities, each coupled to an $^{123}$Sb atom. The two antimony donors share a common electron, which restricts cavity operation to one at a time - two cavity cannot be operated at the same time. 
Comparable coupling strengths, $g$ in Equation~\ref{eq:two_cavity}, (on the order of MHz) can still be achieved via the electric dipole, as the coupling depends on the slope of the hyperfine interaction with respect to the applied electric field, rather than its absolute magnitude. Consequently, the effective coupling to each microwave cavity is expected to be approximately equal.  We select the same EDSR transition (specifically the transitions in between state $\ket{7/2}\ket{\downarrow} \leftrightarrow \ket{5/2}\ket{\uparrow}$, see Fig.~\ref{fig:two_cavity}) for cavity operation. The antimony donors are assumed positioned at an antinode of the cavity
electric field, to achieve capacitive coupling to the donor’s electric dipole and stimulate the EDSR transitions to emit photons coherently.

The total Hamiltonian is given by:
\begin{equation}
    H_{\text{total}} = H_{\text{Sb}_{2}^+} + H_{JC}
\end{equation} where\[ H_{JC} = H_{\text{field}} + H_{\text{interaction}}.\] More precisely,
\begin{equation}\label{eq:two_cavity}
H_{JC}
= \sum_{j \in \{w,s\}}
\left[
\hbar \omega_j\, a_j^{\dagger} a_j
+ g \left(
\sigma^{(j)}_{+} a_j^{\dagger}
+ \sigma^{(j)}_{-} a_j
\right)
\right],
\end{equation}

where \(\omega_j = \omega_{\ket{7/2}_{n_j}\leftrightarrow\ket{5/2}_{n_j}}\) and $
\sigma^{(j)}_{+} = \ket{7/2}_{n_j}\ket{\downarrow}\bra{5/2}_{n_j}\bra{\uparrow}$,
$\sigma^{(j)}_{-} = \ket{5/2}_{n_j}\ket{\uparrow}\bra{7/2}_{n_j}\bra{\downarrow},
$ with \( j \in \{w,s\} \). Subscripts $s$ and $w$ represent the antimony nuclei that are strongly and weakly coupled to the electron respectively.


\subsection{\texorpdfstring{Creation of Qudit Graph States from $\text{Sb}_2^+$}{Creation of Qudit Graph States from Sb2+}}

In this approach, antimony donors are only coupled via a CZ gate when an edge needs to be constructed between the photons, which depends on the shape of the target graph. The protocol therefore requires a CZ gate in addition to all previously established components.

The mutual coupling between nuclear spins is extremely weak. For instance, the magnetic dipole coupling between two nuclei separated by 1 nm is only on the order of 10 Hz. Consequently, most demonstrations of nuclear‑spin entanglement rely on coupling the nuclei to a common electron~\cite{Neumann2008Multipartite, Pfaff2013EntanglementByMeasurement}, requiring nuclear separations of approximately $\approx 1 - 5$ nm, as dictated by the spatial extent of the electron wavefunction~\cite{Abobeih2019AtomicScale, M_dzik_2022}.
Applying a qudit‑CZ gate is therefore not straightforward, since the two antimony nuclei are not directly coupled. However, in the present system—closely related to that in Ref.~\cite{M_dzik_2022}—both nuclei are hyperfine‑coupled to the same electron. By exploiting this shared electron and choosing appropriate nuclear states, one can perform a $2\pi$ rotation of the electron at its ESR frequency, conditional on the two nuclei state, thereby inducing a geometric phase similar to that in Ref.~\cite{Madzik2020}. A potentially more robust approach is to use quantum‑optimal‑control techniques, such as the GRAPE algorithm, to design a CZ gate optimized for the system’s specific Hamiltonian~\cite{omanakuttan, spinor_double_group}.

By coupling the emitters, we can construct arbitrary graph states. However, for convenience, we present two well-known and extensively studied examples from the qubit literature and demonstrate their qudit counterparts: one is a two-dimensional (2D) grid (ladder) graph state, where photons are connected in a 2D lattice similar to matter-based systems; The other example is the well-known 6-ring cluster state, which is commonly used in fault-tolerant surface code implementations within the qubit regime, as it leads to a higher threshold compared to other resource states in this context.~\cite{bartolucci2021fusionbasedquantumcomputation, bombin2023fault, Bombin_2024, bombin2021interleaving}. These examples illustrate how graph states can be deterministically constructed using coupled antimony donors. Note that while these graph states are well-established in the qubit regime, their utility in high-dimensional systems remains an open question. For example, we cannot yet claim that the qudit 6-ring graph state is the counterpart of the qudit surface code, while its qubit version is used for the qubit surface code.  
For now, we simply demonstrate that a wide variety of graph states can be created—and these two serve as representative examples.

\subsubsection{6-Ring Graph state}

To create a 6-ring resource state, we start by applying a CZ gate between two emitters after the first Hadamard gate on the chosen initial states. This entangles the two emitters, thereby creating an edge between the subsequently emitted photons. We then continue with emitting four photons sequentially. After the first CZ gate and four rounds of photon emission, we obtain a linear cluster state of the form $E$–$P$–$P$–$P$–$P$–$E$, where $E$ and $P$ represent the emitter and photons respectively. A second CZ gate between the emitters then closes the structure into a ring. Final emissions and measurements replace the emitters with photons, completing the 6-ring graph state. The protocol for creating qudit 6-ring graph states can therefore be summarized as follows:

\onecolumngrid
\begin{protocol}\label{pro:6-ring}
Deterministic generation of a 6-ring qudit graph state by coupling antimony donors via CZ gates.

\begin{center}
\begin{tikzpicture}[
    node distance=7mm,
    every node/.style={font=\scriptsize},
    process/.style={
        rectangle,
        rounded corners,
        draw,
        text width=8cm,
        align=left,
        minimum height=5mm
    },
    decision/.style={
        diamond,
        draw,
        aspect=2,
        text width=4cm,
        align=center,
        inner sep=1pt
    },
    arrow/.style={->, thick}
]

\node (s1) [process]
{Initialize the system, e.g.,
$\ket{7/2}_{n_s}\ket{7/2}_{n_w}\ket{\downarrow}\ket{\mathrm{vac}}$.};

\node (s2) [process, below=of s1]
{Set photon counter $k=0$.};

\node (s3) [process, below=of s2]
{Apply the qudit Hadamard gate $F_d$ to both antimony nuclei.};

\node (s4) [process, below=of s3]
{Apply a CZ gate between the two antimony nuclei
\textbf{if $k = 0$ or $k = 4$}.};

\node (s5) [process, below=of s4]
{Generate two photons using the steps highlighted in pink in Protocol~\ref{pro:linear_qudit_graph}:
first emitter emits photon $k{+}1$ while second one is idle, followed by
second one emitting photon $k{+}2$ while $n_s$ is idle.};

\node (s6) [process, below=of s5]
{Update the photon counter $k \leftarrow k + 2$ and apply $F_d$ to both nuclei.};

\node (d1) [decision, below=of s6]
{Is $k < 6$?};

\node (s7) [process, below=10mm of d1]
{Measure both antimony nuclei to decouple the nuclear and photonic degrees of freedom.};

\draw [arrow] (s1) -- (s2);
\draw [arrow] (s2) -- (s3);
\draw [arrow] (s3) -- (s4);
\draw [arrow] (s4) -- (s5);
\draw [arrow] (s5) -- (s6);
\draw [arrow] (s6) -- (d1);

\draw [arrow] (d1.east) -- ++(23mm,0) node[midway, above]{Yes}
|- (s4.east);

\draw [arrow] (d1.south) -- node[right]{No} (s7.north);

\end{tikzpicture}
\end{center}
\end{protocol}

\twocolumngrid
\subsubsection{Ladder Graph State}
To illustrate how an arbitrarily shaped graph can be generated by coupling donors without using fusion, we present an additional example: which is a \(2 \times 3\) grid structure, which requires a total of three CZ gates to create the vertical edges. From each emitter, we emit a three-photon qudit linear graph state, and vertical connections are established by applying qudit CZ gates between the two antimony donors.

\begin{protocol}\label{pro:ladder}
Deterministic generation of $2\times3$ Ladder graph by coupling two antimony donors via (qudit) CZ gate.

\begin{center}
\begin{tikzpicture}[
    node distance=6mm,
    every node/.style={font=\scriptsize},
    process/.style={
        rectangle,
        rounded corners,
        draw,
        text width=8cm,
        align=left,
        minimum height=5mm
    },
    arrow/.style={->, thick}
]

\node (s1) [process]
{Choose an initial state for the system. This could be
$\ket{7/2}\ket{7/2}\ket{\downarrow}\ket{\text{vac}}$.};

\node (s2) [process, below=of s1]
{Apply the Fourier gate (qudit Hadamard gate) $F_d$ to both antimony nuclei.};

\node (s3) [process, below=of s2]
{Apply a CZ gate between the antimony nuclei.};

\node (s4) [process, below=of s3]
{Apply the steps highlighted in pink in Protocol~\ref{pro:linear_qudit_graph} to one emitter to generate the first photon while the other remains idle. Then repeat the process with the roles reversed to generate the second photon.};

\node (s5) [process, below=of s4]
{Apply the qudit Hadamard gate $F_d$ to both antimony nuclei.};

\node (s6) [process, below=of s5]
{Apply the steps highlighted in pink in Protocol~\ref{pro:linear_qudit_graph} to one emitter to generate the third photon while the other remains idle. Then repeat the process with the roles reversed to generate the forth photon.};

\node (s7) [process, below=of s6]
{Apply a CZ gate between the antimony nuclei};

\node (s8) [process, below=of s7]
{Apply the qudit Hadamard gate $F_d$ to both antimony nuclei.};

\node (s9) [process, below=of s8]
{Apply a CZ gate between the antimony nuclei};

\node (s10) [process, below=of s9]
{Apply the steps highlighted in pink in Protocol~\ref{pro:linear_qudit_graph} to one emitter to generate the fifth photon while the other remains idle. Then repeat the process with the roles reversed to generate the sixth photon.};

\node (s11) [process, below=of s10]
{Measure both of the nuclei};

\draw [arrow] (s1) -- (s2);
\draw [arrow] (s2) -- (s3);
\draw [arrow] (s3) -- (s4);
\draw [arrow] (s4) -- (s5);
\draw [arrow] (s5) -- (s6);
\draw [arrow] (s6) -- (s7);
\draw [arrow] (s7) -- (s8);
\draw [arrow] (s8) -- (s9);
\draw [arrow] (s9) -- (s10);
\draw [arrow] (s10) -- (s11);

\end{tikzpicture}
\end{center}
\end{protocol}

\begin{figure}[htb]
    \centering
    \includegraphics[width=.8\linewidth]{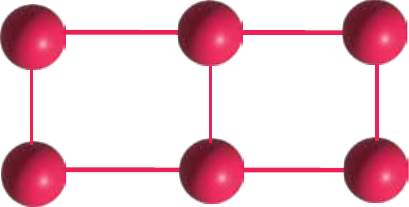}
    \caption[$2\times3$ Ladder Graph]{Visualisation of a $2\times3$ ladder graph. Each vertical edge is created by applying a CZ gate between the two antimony nuclei.}
    \label{fig:ladder}
\end{figure}
\begin{figure*}[htb]
    \includegraphics[width=\linewidth]{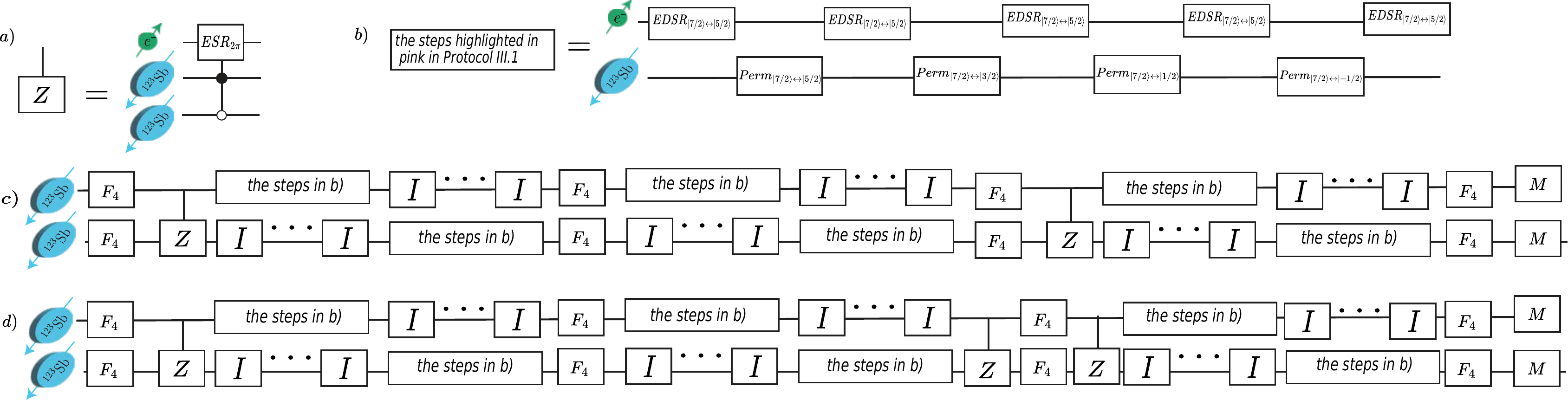}
    \label{fig:6ring}
    \caption{Creation of 6-ring and ladder graph states by coupling antimony donors via CZ gates for qudit dimension $d=4$: (a) illustrates how a CZ gate can be applied between two antimony donors; (b) explicitly shows steps hihglighted in pink in Protocol~\ref{pro:linear_qudit_graph} for four timebins; (c) presents the circuit decomposition for creating a 6-ring graph state by coupling two antimony donors; and (d) shows the circuit decomposition for a 2D ladder graph using coupled antimony donors.}
\end{figure*}

\section{Qudit graphs from Coupled Emitters vs from Single
Emitters}\label{comparison}

The operations involved in creating linear qudit graph states and generating resource states through subsequent fusion operations—along with their experimentally measured fidelities, timescales, and parameters affected by spin-photon interactions and cavity quality factors—are summarized in Table~\ref{tab:qudit_operations} and Table~\ref{tab:photon_loss}. 

For antimony, the predominant error source is phase flip, though operation errors remain very low (Table~\ref{tab:qudit_operations}). ESR (EDSR) pulses takes $1~\mu$s ($10~\mu$s). A qudit Hadamard takes up to $100~\mu$s, while permutations may take a few hundred $\mu$s. The electron $T_2^*$ is $510~\mu$s and the nuclear $T_2^H$ is $247~\mu$s~\cite{fernandez2024navigating}. Time-bin multiplexing enables general superpositions within both $\ket{\downarrow}$ and $\ket{\uparrow}$ manifolds, allowing access to more than eight levels, limited by antimony coherence times. Hardware advances, including improved $^{28}$Si enrichment~\cite{acharya2024highly}, enhanced readout~\cite{santiagosito}, and novel microwave antennas~\cite{dehollain2012nanoscale}, aim to extend coherence times.

Photon emission is a coherent process governed by the Jaynes-Cummings model, and the time ($t_e$) required to emit a photon is dependent upon the coupling strength. $t_e = 1/g_s$, with strong coupling through electric dipole $g_s = 3$~MHz~\cite{tosi_2017} which corresponds to $g$ in Equations~(\ref{drift_Ham},\ref{eq:two_cavity}), giving $t_e \approx 333/2\pi$~ns. The electron $T_1$ is several seconds ($2.44$~s~\cite{fernandez2024navigating}), ensuring the system does not decay to other states during emission. Spin dephasing occurs at $1/T_2^*$, with $T_2^* = 510~\mu$s for $m_I = \ket{7/2}$~\cite{fernandez2024navigating}. Thus, photon generation rate ($3$~MHz) is three orders of magnitude faster than electron dephasing ($1.96$~kHz).  

Photon loss due to the cavity quality factor $Q$ involves two components for microwave cavities: (1) $Q_i$: the internal quality factor of the cavity, which characterizes the rate at which photons are lost to the bath. Based on our previous findings \cite{vaartjes2023strong, Wyatt_1, wyatt_2}, this can be taken to be in the range $10^5$ to $10^6$. (2) $Q_c$: the coupling quality factor, which characterizes the rate at which photons go to the detectors, $10^4$~\cite{vaartjes2023strong}. The coupling rates are $\kappa_i = \omega_c/Q_i$ and $\kappa_c = \omega_c/Q_c$. With $g_s = 3$~MHz~\cite{tosi_2017}, the effective rates are $ \gamma_{\text{bath}} = \frac{g_s \kappa_i}{g_s + \kappa_i}, \quad 
\gamma_{\text{port}} = \frac{g_s \kappa_c}{g_s + \kappa_c}.$
Photon loss is $\text{Loss} = \frac{\gamma_{\text{bath}}}{\gamma_{\text{bath}} + \gamma_{\text{port}}},$
while the detection probability is
$ \text{Success} = \frac{\gamma_{\text{port}}}{\gamma_{\text{bath}} + \gamma_{\text{port}}}, $
with success in dB given by $ \text{dB} = 10 \log_{10}(\text{Success}).$
The relevant EDSR transition, $\ket{7/2}\ket{\downarrow} \leftrightarrow \ket{5/2}\ket{\uparrow}$, occurs at $28.41$~GHz~\cite{fernandez2024navigating}, so we take $\omega_c = 28.41$~GHz for cavity resonance.

\begin{table*}[htb]
\centering
\begin{tabular}{|c|c|c|}
\hline
\textbf{Operation} & \textbf{Fidelity $\%$} & \textbf{Time Scale} \\
\hline
State Initialization & 99.5 & a few tens of ms \\
\hline
NMR pulses (or single qubit nuclear gate) & 99.8 & ~50 $\mu$s\\
\hline
ESR pulses & 99.5 & 1 $\mu$s \\
\hline
EDSR pulses & 99.5 &  8.5 $\mu$s\\
\hline
Largest permutation gate (between state $\ket{-7/2}$ and $\ket{7/2}$)& 91.5 &  ~500 $\mu$s\\
\hline
Measuring single donor & $>$ 99 &  10 - 100 ms\\
\hline
$T_1$ for electron & N/A & 2.44 s\\
\hline
$T_2^*$ for electron & N/A & 11.06 $\mu$s\\
\hline
$T_2^H$ for electron & N/A & 510 $\mu$s \\
\hline
$T_2^H$ for antimony & N/A & 247 $\mu$s\\
\hline
$T_2^*$ for antimony & N/A & a few $\mu$s $^*$\\
\hline
\end{tabular}
\caption{Summary of key operations involved in linear qudit graph state generation, along with measured timescales and fidelities. The values are taken from~\cite{Yu2025,fernandez2024navigating}. Note that the $T_2^*$ for antimony was not measured in Ref~\cite{Yu2025, fernandez2024navigating}.}
\label{tab:qudit_operations}
\end{table*}

\begin{table*}[htb]
\small 
\centering
\begin{tabularx}{\textwidth}{|l|X|X|}
\hline
\textbf{Parameter} & \textbf{Value / Formula} & \textbf{Notes / Source}  \\
\hline
\textcolor{red}{Cavity frequency (\(\omega_c\))} & \textcolor{red}{\(28.41\,\text{GHz}\)} & \textcolor{red}{EDSR transition frequency} \\
\hline
Spin-photon coupling strength (\(g_s\)) & \(3\,\text{MHz}\) & From~\cite{tosi_2017}  \\
\hline
Internal quality factor (\(Q_i\)) & \(10^5 - 10^6\) & Photon loss to bath~\cite{vaartjes2023strong, Wyatt_1, wyatt_2} \\
\hline
Coupling quality factor (\(Q_c\)) & \(10^4\) & Photon coupling to detector~\cite{vaartjes2023strong} \\
\hline
\(\kappa_i\) (bath coupling rate) & \(\omega_c / Q_i\) & Cavity-to-bath rate \\
\hline
\(\kappa_c\) (detector coupling rate) & \(\omega_c / Q_c\) & Cavity-to-detector rate  \\
\hline
\(\gamma_{\text{bath}}\) & \(\frac{g_s \kappa_i}{g_s + \kappa_i}\) & Effective photon loss rate  \\
\hline
\(\gamma_{\text{port}}\) & \(\frac{g_s \kappa_c}{g_s + \kappa_c}\) & Effective detection rate  \\
\hline
Photon loss ratio & \(\frac{\gamma_{\text{bath}}}{\gamma_{\text{bath}} + \gamma_{\text{port}}}\) & Denoted as \texttt{Loss} \\
\hline
Photon success ratio & \(\frac{\gamma_{\text{port}}}{\gamma_{\text{bath}} + \gamma_{\text{port}}}\) & Denoted as \texttt{Success}  \\
\hline
Success (in dB) & \(10 \log_{10}(\text{Success})\) & Photon transmission efficiency  \\
\hline
\multicolumn{3}{|c|}{\textbf{Example Values}} \\
\hline
\(Q_i = 10^6\) & Loss = 0.0189 & Success = 0.981 \(0.08\,\text{dB}\) \\
\hline
\(Q_i = 10^5\) & Loss = 0.15 & Success = 0.845 \(0.7\,\text{dB}\) \\
\hline
\textbf{Simulated photon loss range} & 1\%–10\% per mode  & Best- and worst-case scenarios \\
\hline
\end{tabularx}
\caption{Summary of cavity-related parameters, photon loss and success probabilities, and their role in qudit graph state generation. The cavity frequency, highlighted in red, will be different for the Sb$_2$ molecule since the hyperfine interaction strengths are different. However, as noted above, we can still achieve the similar coupling strength (orders of a few MHz), since it depends not on the magnitude of the hyperfine interaction but on its slope with respect to the electric field.}
\label{tab:photon_loss}
\end{table*}

In addition to the operations within the antimony donor and photon emissions through microwave cavity, we also consider fusion operations performed on the produced graph states, along with their success probabilities. The success probability is given by \( \frac{2}{d(d + 1)} \) for odd dimensions and \( \frac{2}{d^2} \) for even dimensions, which corresponds to 0.16 for \(d = 3\) and 0.055 for \(d = 4\).

For the Sb$_2$ molecule, several parameters will differ from those of a single antimony donor. Since the device is different, all coherence times will also be different. The measurement (readout) time will also vary, and the readout fidelity will be lower than in the single-antimony donor. However, the timings and associated error rates for ESR, EDSR, and NMR (primitive pulses) are expected to remain approximately the same. The photon generation rate will still be orders of magnitude greater than the dephasing rate of antimony, and the $T_1$ time is expected to be on the order of seconds, ensuring that the system does not decay to other states during emission, as in the single-antimony-donor case 

In the case of coupled donors, the creation of a single resource state is not affected by the low success probability of fusion operations because no fusion is required. By coupling two antimony donors with a CZ gate, we eliminate the need for probabilistic fusion in generating resource states. The CZ gate, executed via ESR pulses, takes only a few microseconds, enabling (in principle) deterministic fusion. We are, however, bound by the coherence time of the Sb$_2$ molecule. 

For a more concrete comparison, let us consider the creation of a six-node ring graph state for qudits of dimension three under these two scenarios:  In terms of generation time, the preparation of a qutrit single-photon remains the same in both cases, as the durations for the EDSR pulse (8.5~$\mu$s), the qudit Hadamard pulse (applied to the single antimony donor in both scenarios, typically 100~$\mu$s), and the permutation operations are identical. For the qutrit case, permutation operations between $\ket{7/2}$ and $\ket{5/2}$ ($\approx 50~\mu$s), $\ket{5/2}$ and $\ket{3/2}$ ($\approx 100~\mu$s), and $\ket{3/2}$ and $\ket{1/2}$ ($\approx 150~\mu$s) result in a total of about 300~$\mu$s without optimal quantum control techniques. Additionally, three EDSR transitions are required for emission into three time bins, totaling $8.5 \times 3 = 25.5~\mu$s. Operations on the antimony donor take a few hundred microseconds, while photon emission from the microwave cavity occurs on the nanosecond scale. The total time required to create a qutrit single-photon is therefore $100 + 300 + 25.5 \approx 425.5 \mu$s. For the single‑emitter case, this process must be repeated eight times ($(425.5 * 8) \mu$s in total) before the donor measurement, because two of the photons will be used in the Type‑II fusion. For the coupled‑emitter case, six repetitions are required ($(425.5 * 6) \mu$s in total), followed by the donor measurement.

Furthermore, the single-emitter approach combined with fusion requires multi-dimensional Fourier projections (hundreds of nanoseconds) and suffers from the non-deterministic nature of fusion operations. For the qutrit case, the success probability of fusion is only 0.16. In contrast, qudit CZ gates between donors are deterministic and require only 1--2~$\mu$s each, with two qudit CZ gates needed in total for the 6-ring generation. While fusion operations occur on the nanosecond scale, their success probability is extremely low for qudits and requires an entangled ancilla state, which may involve additional preparation using antimony donors depending on the dimension. In addition to this, in the case of unsuccessful fusion—which is highly likely in the qudit regime—we must repeat the fusion process and reprepare both the ancilla and the photons involved in the fusion until we successfully create the target resource state.  Therefore, coupled emitters with deterministic CZ gates outperform single emitters combined with fusion operations in terms of both overhead and overall success probability for generating qudit graph states. This arises primarily from the low success probability of fusion operations in the qudit setting.

One challenge with the Sb$_2$ molecule is the presence of two microwave cavities when the donors are only 5 nm apart: in principle, we can fabricate two separate cavities; however, there is a significant possibility that both cavities would couple to both donors, as it is difficult to localise the microwave field sufficiently when the donors are only 5 nm apart. Nevertheless, because the resonators would be tuned to differ by approximately 95 MHz, independent coupling and operation should still be achievable.
This 95 MHz separation between the EDSR frequencies, however, introduces a new issue: it can make the emitted photons distinguishable. Although similar coupling strengths—on the order of a few MHz—can be achieved for both, they operate at different frequencies ( highlighted in red in the table) because one antimony donor is weakly coupled to the electron and the other is strongly coupled. The corresponding hyperfine interaction strengths are 96 MHz and 239 kHz, resulting in a cavity frequency difference of about 95 MHz. The EDSR frequency is given by $B_0 \gamma_+ + (m_I - \frac{1}{2})(f_q + A)$, where $\gamma_+ = \gamma_n + \gamma_e$, $f_q$ is the quadrupole splitting frequency, which is on the order of a few tens of kilohertz and $A$ is the hyperfine interaction strength. This frequency difference does not affect cavity performance, as both still operate in the GHz range, similar to the single-donor case. However, the emitted photons will have different frequencies and will therefore be distinguishable. The constraint arises at the architectural level, where fusion between multiple and/or different resource states requires frequency matching. Fusion cannot be performed between photons of different frequencies. This limitation can be addressed by designing the architecture to fuse only frequency-matched photons or by scheduling photon generation so that photons intended for fusion are produced consecutively and positioned adjacently in space or time. This ensures compatibility for fusion operations. Nonetheless, generating identical photons remains preferable.

The table summarizes the values that differ from those of the single antimony donor device can be found in~\ref{tab:qudit_operations2}.

\begin{table*}[htb]
\centering
\begin{tabular}{|c|c|c|}
\hline
\textbf{Operation} & \textbf{Fidelity $\%$} & \textbf{Time Scale} \\
\hline
CZ gate between the two donors & not yet benchmarked & a few $\mu$s \\
\hline
Measuring $Sb_2$ molecule &  90 &  1 ms\\
\hline
$T_1$ for electron & N/A & not yet measured \\
\hline
$T_2$ for electron & N/A &  6.3$\mu$s\\
\hline
$T_2$ for antimony donors & N/A &  2 $\mu$s\\
\hline
$T_2^*$ for electron & N/A & not yet measured \\
\hline
$T_2^*$ for antimony & N/A & not yet measured\\
\hline
$T_2^H$ for electron & N/A & not yet measured \\
\hline
$T_2^H$ for antimony & N/A & not yet measured\\
\hline
\end{tabular}
\caption{Summary of key operations and coherence properties that differ from the single-donor device for creating arbitrary-shaped graph states using the Sb$_2$ molecule. The values are taken from~\cite{Hsu2025}.
}
\label{tab:qudit_operations2}
\end{table*}

With the Sb$_2$ molecule, another interesting state that can be generated—without the need for fusion operations—is the absolutely maximally entangled (AME) state. AME states are a special class of multipartite entangled states that are maximally entangled for any possible bipartition~\cite{helwig2013absolutelymaximallyentangledqudit}. 
In our example, the system is capable of producing such a state with four photons, each having a qudit dimension of 3. More detail and example of four qutrit AME states can be found in~\cite{Raissi_2024}.

\section{Sources of Device-to-Device Variation in Antimony Silicon Chips}
As discussed in previous sections, even within the same device, the hyperfine interactions, quadrupole terms, and ESR/EDSR transitions can vary depending on the exact location where the donor ends up implanted on the silicon surface.
This raises an important question: how similar can antimony-based silicon chips be, and what factors contribute to variations in their characteristics?

Potential sources of variation in ESR frequencies across devices:
\begin{itemize}
    \item \textbf{The static $B_0$ field}: Devices may operate under different static magnetic fields $B_0$, which directly influence the Zeeman splitting of the electron and, consequently, its resonance frequency. The choice of magnetic field can depend on whether the device is placed in a magnet box or within a solenoid magnet. However, if both devices are subjected to the same magnetic field, they should ideally exhibit identical Zeeman splitting. This is not a concern for a device with $N$ implanted donors, since all donors within the same device experience the same $B_0$ field.

    \item \textbf{The hyperfine interaction}: The hyperfine coupling can vary depending on the donor's position relative to the gates and the specific gate architecture, which may differ slightly between devices due to nanofabrication imperfections. These variations can induce strain on the donor, thereby modifying the hyperfine interaction. Typically, such changes are on the order of a few MHz ($\sim <$ 5~MHz), although larger deviations may occur, particularly when the donor is located near the interface and is significantly affected by gate-induced strain. 
\end{itemize}

Potential sources of variation in neutral NMR frequencies across devices:
\begin{itemize}
    \item \textbf{The $B_0$ field}: As previously noted, variations in the static magnetic field affect the nuclear Zeeman splitting.

    \item \textbf{The hyperfine interaction}: As described above, this depends on the donor's location and the specific gate configuration.

    \item \textbf{The quadrupole interaction}: This term is also strain-dependent and varies with the donor's position within the device and the surrounding gate structure. Observed quadrupole values range from approximately 4~kHz to 50~kHz, indicating typical variations on the order of tens of kHz, although the current dataset remains limited. 
\end{itemize}

Potential sources of variation in ionised NMR frequencies across devices:
\begin{itemize}
    \item \textbf{The $B_0$ field}: As previously discussed, this influences the Zeeman splitting.

    \item \textbf{The quadrupole interaction}: As noted above, this term is sensitive to the strain experienced by the donor.
\end{itemize}

Potential sources of variation in EDSR frequencies across devices:
\begin{itemize}
    \item \textbf{The $B_0$ field}: As mentioned earlier, this affects the resonance conditions.

    \item \textbf{The hyperfine interaction}: As previously described, this is strain-dependent.

    \item \textbf{The quadrupole interaction}: As discussed above, this also varies with strain.
\end{itemize}

Note that even within the same device, donors can experience small variations in their frequencies due to variations in the hyperfine interaction and the quadrupole interaction.
\section{Conclusion - Outlook}
We presented a comparative study on creating qudit cluster states. One approach is based on creating linear qudit graph states from a single antimony donor. These states are then fused using a novel technique introduced in~\cite{Ustun_fusion} to create resource states—graph states more complex than linear ones.

Subsequently, we introduced a two-antimony, one-electron system, where both antimony donors share a common electron. This system is realized as a molecule, and the donors are not implanted individually using ion implantation. As a result, upon implantation, the two antimony atoms are approximately 5 nm apart. One donor is strongly coupled to the shared electron, while the other is weakly coupled. This asymmetry leads to a significant difference in hyperfine interactions: the donor strongly coupled to the electron exhibits a hyperfine interaction that is 95 MHz higher than that of the weakly coupled donor.

Using the Sb$_2$ system, we propose employing two microwave cavities—one for each antimony donor—to directly create arbitrary qudit graph states, and we provide examples for a 6-ring and a 2D ladder graph state. This approach eliminates the need for fusion methods by applying a CZ gate between the two donors, effectively replacing fusion with a deterministic CZ gate.

However, this method introduces several challenges. First, due to the molecular implantation rather than individual ion implantation, the shared electron couples asymmetrically to the donors, resulting in significantly different hyperfine interactions. Since we propose using the cavity on the EDSR transition, the operating frequencies of the cavities differ by 95 MHz, making the emitted photons distinguishable.

Additionally, because the molecule uses a single shared electron, only one photon can be emitted at a time. This leads to long circuit operation times, with one of the nuclei often idling.
To overcome this, we propose implanting the donors individually, placing them $\approx 20 $ nm apart instead of 5 nm. In this configuration, each antimony donor can have its own electron, allowing simultaneous photon emission and eliminating idle operations in circuits. This approach shortens circuit operation times and ensures that the hyperfine interactions are within the same range, making the emitted photons highly likely to be indistinguishable. 

When it comes to the scale of $N$ donors from the Sb$_2$ molecule, Two important points can be highlighted: 1) molecule implantation is not practical for scalibility; instead, single antimony donors must be implanted. This approach minimizes the 95 MHz hyperfine interaction difference between antimony donors on the same silicon chip to just a few megahertz, which is crucial for emitting indistinguishable photons. 2) Deterministic implantation is essential compared to timed implantation, as it allows precise control over donor positions with negligible uncertainty. Each donor should be at least 20 nm apart, as demonstrated in~\cite{Stemp2024,Holly_science}, ensuring that each donor retains its own electron. This spacing is not only critical for simultaneous photon emission rather than sequential emission, but also facilitates integration—specifically, the fabrication of microwave cavities on individual donors becomes significantly more feasible. When resonators operate at approximately the same frequencies—more precisely, when their frequency difference is less than about a few tens of MHz—they are likely to be strongly coupled, leading to hybridized modes. If the donor separation is very small (e.g., around 5 nm), the entire system effectively couples together, and independent control and operation of the cavities are lost. In contrast, when the distance between donors is on the order of 50 nm, independent control and operation of the resonators (microwave cavities) become feasible, even when the cavity frequencies are close to one another.

The core challenge for scalability to $N$ (where $N$ is typically much smaller than the number of qubits in solid-state hardware) lies in fabrication advancements. Once deterministic ion implantation of single donors becomes standard in laboratories, the scalability limitations associated with Sb$_2$ molecules will be largely mitigated. In addition, if a better fusion method can be found whose success probability surpasses that of the qubit case, this would in itself be a significant advantage. In such a scenario, we could operate with silicon chips containing single emitters and fabricate many of them under almost identical conditions using the techniques mentioned above. This would remove the necessity of implanting many resonators coupled to multiple donors within the same device. However, at present, we do not have such fusion operations. This is what makes coupled emitters more advantageous than a single emitter combined with fusion operations, despite their fabrication difficulties in the qudit regime; in the coupled-emitter case, we do not require fusion, at least at the first step, and can deterministically create the targeted resource state.

For future work, two distinct directions can be pursued: experimental and theoretical. On the experimental side, realizing a system in which a single antimony donor is coupled to a microwave cavity is particularly exciting. Building on this, the next step would be to fabricate a system with two microwave cavities coupled to two antimony donors. As mentioned above, implanting individual antimony donors rather than implanting them as molecules offers several advantages. From a theoretical perspective, a key objective is to explore FBQC in higher dimensions beyond the physical ingredients. This includes understanding which resource states are best suited for different quantum error correction codes, as well as establishing fusion-based quantum computing for qudits, which requires analyzing whether the success probabilities of fusion operation in high-dimensional systems~\cite{Ustun_fusion} are sufficient.

\section{Acknowledgments}
We thank Samuel Elman, Wyatt Wine, Jason Saied, Benjamin Wilhelm, Jarryd Pla, Andrew Doherty for stimulating discussions throughout the project. We thank Sean Hsu for providing us Hamiltonian paramaters for coupled antimony. G.\"U. acknowledges support from Sydney Quantum Academy (SQA) where she is a primary scholarship holder. G.Ü. acknowledges the support of the Australian Research Council (ARC) Center of Excellence in Quantum Computing and Communications Technology (CQC2T). Project led by University of Technology Sydney and supported by Defence Science and Technologies Group (DSTG) and Advanced Strategic Capabilities Accelerator (ASCA) through its Emerging and Disruptive Technologies (EDT) Program.
\newpage
\section*{Appendix}
\setcounter{section}{0}
\setcounter{equation}{0}
\setcounter{figure}{0}
\renewcommand{\thesection}{A-\Roman{section}}
\renewcommand{\theequation}{A.\arabic{equation}}
\renewcommand{\thefigure}{A.\arabic{figure}}

\section{Linear Qubit graph State from Single Antimony}

Here, we show how linear qubit graph states can be created from a single antimony donor.

\begin{figure}[htb]
    \centering
    \includegraphics[width=.8\linewidth]{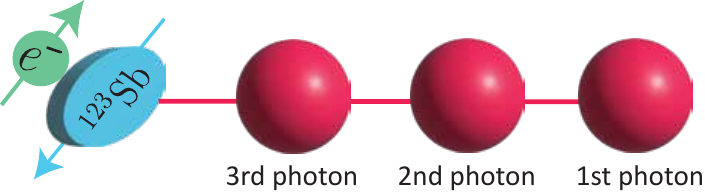}
    \caption{A three-qubit linear graph state is deterministically generated from antimony using time-bin multiplexing scheme. At the end of the graph state generation, antimony donor is measured to decouple it from the photons. As a result, the final state consists solely of photonic qubits.}
    \label{fig:qubit_graph}
\end{figure}


By selecting two of the eight ground states of antimony and the corresponding excited state associated with one of these ground states, where the cavity operates at the EDSR transition, we can generate linear qubit graph states. 

The protocol starts with the initial state $\ket{\psi_0} = \ket{7/2}\ket{\downarrow}\ket{vac}$, and by applying a Hadamard operation on the two nuclear spin states of antimony - just like the example above - the resultant states becomes:
\[
    \begin{split}
        \ket{\psi_1}=\frac{1}{\sqrt{2}}\big( \ket{7/2} + \ket{5 /2} \big) \ket{\downarrow}\ket{vac},
    \end{split}
\]
where the electron is in the spin-down state $\ket{\downarrow}$. Then an EDSR pulse is applied, which flips the spin of the electron conditioned on the nucleus being in the state $\ket{7/2}$:
\[
    \begin{split}
        \ket{\psi_2}=\frac{1}{\sqrt{2}} \ket{7/2}\ket{\uparrow}\ket{vac} + \frac{1}{\sqrt{2}}\ket{5 /2} \ket{\downarrow}\ket{vac}
    \end{split}
\]
The electron experiences a coherent exchange of energy with the microwave cavity such that at a time $t_1$ 
the electron is in the down state and the cavity is populated with a photon of frequency ${\omega}$:
\[
    \begin{split}
        \ket{\psi_3}=\frac{1}{\sqrt{2}} \ket{7/2}\ket{\downarrow}\ket{\omega}_{t1} + \frac{1}{\sqrt{2}}\ket{5 /2} \ket{\downarrow}\ket{vac}
    \end{split}
\]
We apply an NMR $\pi$ pulse between the $\ket{7/2}$ and $\ket{5/2}$ states to exchange their populations, transferring the population from $\ket{5/2}$ to $\ket{7/2}$. Then, using an EDSR pulse and the cavity, we attempt to emit a photon. The resulting state is:
\[
    \begin{split}
        \ket{\psi_4}=\frac{1}{\sqrt{2}} \ket{7/2}\ket{\downarrow}\ket{\omega}_{t1}\ket{vac} + \frac{1}{\sqrt{2}}\ket{5 /2} \ket{\downarrow}\ket{vac}\ket{\omega}_{t2}
    \end{split}
\]
A single photon is emitted into two time-bins where $\ket{\omega}_{t1}\ket{vac} = \ket{10} = 1$ and $\ket{vac}\ket{\omega}_{t2} =\ket{01} = 0 $ being the encoding parameters. 

We now repeat the entire procedure, starting with the application of a second Hadamard gate on the spin states of the antimony.
\[ 
    \begin{split}
        \ket{\psi}_{\text{2nd cycle}}=\\&\frac{1}{{2}} (\ket{7/2}\ket{\omega}_{t1}\ket{vac} + \ket{5/2}\ket{vac}\ket{\omega}_{t2}) \\&\ket{\downarrow}\ket{\omega}_{t1}\ket{vac} +\\& \frac{1}{{2}}(-\ket{5 /2}\ket{vac}\ket{\omega}_{t2} +\ket{7/2}\ket{\omega}_{t1}\ket{vac}) \\&\ket{\downarrow}\ket{vac}\ket{\omega}_{t2}
    \end{split}
\]
Applying the final Hadamard operation, after which the antimony is measured to decouple the spin from the two photons, and the resulting two-photon state takes the following form when expressed in binary:
\[
    \begin{split}
       \ket{\psi_7}=  &\frac{1}{2\sqrt{2}}(\ket{7/2}-\ket{5/2})(\ket{10}+\ket{00})+ \\&\frac{1}{2\sqrt{2}}(\ket{7/2}+\ket{5/2})(\ket{11}-\ket{01})\\
       =&\frac{1}{2\sqrt{2}}\ket{7/2}(\ket{10}+\ket{00} + \ket{11}-\ket{01})+ \\ & \frac{1}{2\sqrt{2}}\ket{5/2}( \ket{11}-\ket{01} - \ket{10}-\ket{00})
    \end{split}
\] 
This is a two-qubit linear graph state, after we measure (readout) the antimony. Alternatively, we could proceed with the emission after the third Hadamard gate and then apply the final Hadamard to decouple the spin from the three photons. This would result in a three-qubit linear graph state, just as we obtained in the previous example involving polarization and quantum dots.
\[
  \begin{split}
       \ket{\psi_{\text{third cycle}}}= 
       &\ket{7/2}(\ket{101}+\ket{001} + \ket{111}-\ket{011} + \\& \ket{110}-\ket{010} - \ket{100}-\ket{000})+\\&\ket{5/2}(\ket{000}+\ket{010}+ \ket{100}-\ket{110}-\\& \ket{101}-\ket{001} - \ket{111}-\ket{011} )
    \end{split}  
\]
As a result, we can generate qubit cluster states using antimony by selecting any two levels from its eight available states. 

\subsection{Resource efficient qubit FBQC using high-dimensional graph states}
One important use case for high-dimensional graph states is their application in encoding qubit cluster states~\cite{Lib_2024}. By leveraging antimony donors, it is possible to encode multi-(qubit)-photons into a single high-dimensional qudit photon—for example, storing three qubits within one photon ($2^3 = 8$-dimensional single photon). It has been demonstrated that a 24-photon resource state is required for fault-tolerant quantum computing in photonic architectures~\cite{bartolucci2021fusionbasedquantumcomputation}. Using the high-dimensional encoding method described above, together with time-bin multiplexing, we can generate eight high-dimensional photons that form a linear graph state, with each photon encoding three qubit photonic states. This approach effectively replaces 24 photonic qubits with eight photonic qudits in a graph configuration, reducing the total number of required photons by a factor of three.

Moreover, linear optical elements such as beam splitters and interferometers can implement arbitrary unitaries within a single qudit, enabling intra-qudit operations that are not straightforward in conventional qubit-based schemes. However, this introduces a significant challenge: photon loss. In photonic quantum computing, photon loss is a dominant error source, and losing a single qudit photon in this encoding scheme results in the loss of multiple qubits' worth of information.

This issue can be addressed using quantum error correction tailored to qudits as we mentioned in the introduction. For example, the $[[5,1,3]]_d$ modular-qudit code~\cite{chau1997five} allows for encoding one logical $d$-dimensional qudit into five physical qudits, tolerating up to two erasures (i.e., photon losses) regardless of the dimension $d$. In essence, using qudits enables the encoding of more information per physical system while maintaining robustness to photon loss, without increasing the number of physical systems required.

\begin{thebibliography}{56}
\bibitem [{spi()}]{spinor_double_group}%
  \BibitemOpen
  \href@noop {} {}\bibinfo {note} {Half-integer spins are known as spinors and belong to the double group in condensed matter physics. A $2\pi$ rotation accumulates a phase and results in a geometric phase gate rather than the identity operation. A second $2\pi$ rotation restores the identity. This behavior does not apply to integer spins.}\BibitemShut {Stop}%
\end{thebibliography}
%

\end{document}